\documentclass{epl}

\title{Quenching of Lamellar Ordering\\ in an n-Alkane Embedded in Nanopores}
 \author{P. Huber \and D. Wallacher \and J. Albers \and K. Knorr}
\institute{Fakult\"at f\"ur Physik und Elektrotechnik,
Universit\"{a}t des Saarlandes, D-66041 Saarbr\"{u}cken, Germany}

\pacs{61.46.+w}{Nanoscale materials: clusters, nanoparticles,
nanotubes, and nanocrystals}

\pacs{64.70.Nd}{Structural transitions in nanoscale materials}

\pacs{61.10.Eq}{X-ray scattering (including small-angle
scattering)}

\begin{document}

\maketitle

\begin{abstract}
We present an x-ray diffraction study of the normal alkane
nonadecane $C_{19}H_{40}$ embedded in nanoporous Vycor glass. The
confined molecular crystal accomplishes a close-packed structure
by alignment of the rod-like molecules parallel to the pore axis
while sacrificing one basic ordering principle known from the bulk
state, i.e. the lamellar ordering of the molecules. Despite this
disorder the phase transitions observed in the confined solid
mimic the phase behavior of the 3D unconfined crystal, though
enriched by the appearance of a true rotator phase known only from
longer alkane chains.
\end{abstract}

The goal of manipulating matter on the nanometer scale has led to
a particularly rich area of research on how basic concepts of
condensed matter physics, i.e. geometric ordering principles like
close-packing, phase transition phenomenology and thermodynamics
are changed once one reaches this length scale. In fact, there has
already been a tremendous increase in understanding of such
phenomena by experimental \cite{Christenson2001} and theoretical
\cite{Gelb1999} studies of fluids in nanoporous glasses. These
randomly interconnected porous networks are unique systems in that
they allow us to easily prepare and study systems in restricted
geometry. It has been known for a long time that the freezing
point of liquids is reduced and that the "capillary" condensation
of the vapor occurs at a reduced vapor pressure in such
geometries\cite{Christenson2001}. Little however is known of the
impact of confinement on the solid state of matter. The few pore
solids investigated so far had globular building blocks,
significantly smaller than the mean pore diameter of the porous
networks. They show structures closely related to the bulk state.
Prominent examples are krypton, argon, nitrogen, oxygen and
H$_{2}$0\cite{Morishige2000Krypton,Huber1999,Wallacher2001,Morishige1999H2O}.

In this Letter, we focus on the structure and phase transitions of
a more complex molecular crystal, built of chain-like molecules
whose lengths are comparable to the mean diameter of the pores.
Our temperature dependent x-ray study elucidates how these
confined molecules establish one of the basic building principles
of condensed matter, i.e. close packing, despite drastic geometric
restrictions in the pores. Furthermore, an analysis of the
temperature dependent lattice parameters reveals that the confined
alkane mimics order-disorder transitions known from the
three-dimensional (3D) unconfined crystals.

\begin{figure}
\onefigure{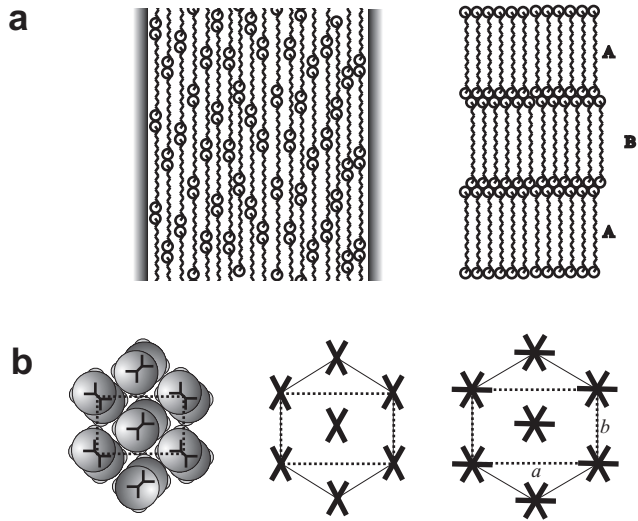} \caption{(a)~A schematic side view of
n-alkane C$_{19}$H$_{40}$ in a nanopore and of the layered
structures of the bulk phases C and R$_{I }$ with a stacking
scheme AB. (b)~A view along the chain axis of the bulk phases C,
R$_{I}$, R$_{II}$ (from left to right) showing the various degrees
of order and disorder of the orientation of the -C-C- plane.}
\label{fig01}
\end{figure}

For our x-ray powder diffraction study, we have chosen the
n-alkane C$_{19}$H$_{40}$ embedded in Vycor glass. From a
transmission electron microscope photograph\cite{Levitz1991},
Vycor can be described as a network of 3D randomly connected
tubular pores with relatively uniform diameter $d\sim 70\AA$ and
average length $l \sim 300\AA$. The normal alkanes are one of the
most basic organic series; they are the building blocks for
surfactants, liquid crystals and lipids. Crystalline
C$_{19}$H$_{40}$ can be considered an almost rigid rod-like
molecule with a length of about $25\AA$ and a width of $4\AA$. The
C atoms of the zig-zag backbone are all in the trans
configuration, so that all of them are located in a plane. Gauche
defects which lead to kinks and twists of the -C-C- chain are
abundant in longer alkanes, but have little effect on the crystal
structure of the phases of bulk
C$_{19}$H$_{40}$\cite{Craievich1984}. In the bulk state the
medium-length alkanes form layered crystals \cite{Mueller1932}
(see Fig.~\ref{fig01}). For C$_{19}$H$_{40}$ the molecules are
aligned perpendicular to the layers \cite{Sirota1993}. Within the
layers the molecules are 2D close packed, side by side, in a
quasihexagonal 2D array (Fig.~\ref{fig01}), described by in-plane
lattice parameters $a$ and $b$. The azimuth of the rotation of the
-C-C- plane around the long z-axis of the molecule defines an
orientation. In the low-temperature phase of the bulk alkane,
termed "crystalline" C, there are two different orientations which
alternate in a herringbone fashion. Upon heating this phase
undergoes a phase transition at 294.8~K into a state of partial
orientational disorder, the so-called "rotator" phase R$_{I
}$(\cite{Sirota1993}\cite{Doucet1981}). Here the molecules flip
between two equivalent positions. This phase melts at 304.5~K.
Somewhat longer alkanes C$_{n}$H$_{2n + 2}$, $21~< n < 26$,
display another mesophase with complete orientational disorder
\cite{Sirota1993}, the phase R$_{II}$, in which the molecules
perform hindered rotations or flips around the z-direction between
3 equivalent orientations. This phase has a strictly hexagonal
metric, that is $a/b = \surd $3. See Fig.~\ref{fig01} for the
orientational pattern of the three phases C, R$_{I}$, R$_{II}$.
Fig. 2 shows the x-ray powder pattern of the R$_{I}$ phase of the
bulk solid. The dominant peaks are a series of (00l) reflections
at lower Bragg angles $2\theta $, which are related to the layered
arrangement of the molecules, and the two principal in-plane
reflections around 22deg. In the hexagonal lattice of phase
R$_{II}$ these two peaks would merge.

For the study of the pore condensate, the Vycor glass matrix has
been soaked with the melted alkane in an evacuated glass chamber
at $T=310K$. Weight measurements of the glass monolith before and
after filling, the known mass density of the bulk liquid, and the
specific pore volume allowed us to calculate a filling fraction of
$98\pm 2\% $ of the pores.

\begin{figure}
\onefigure{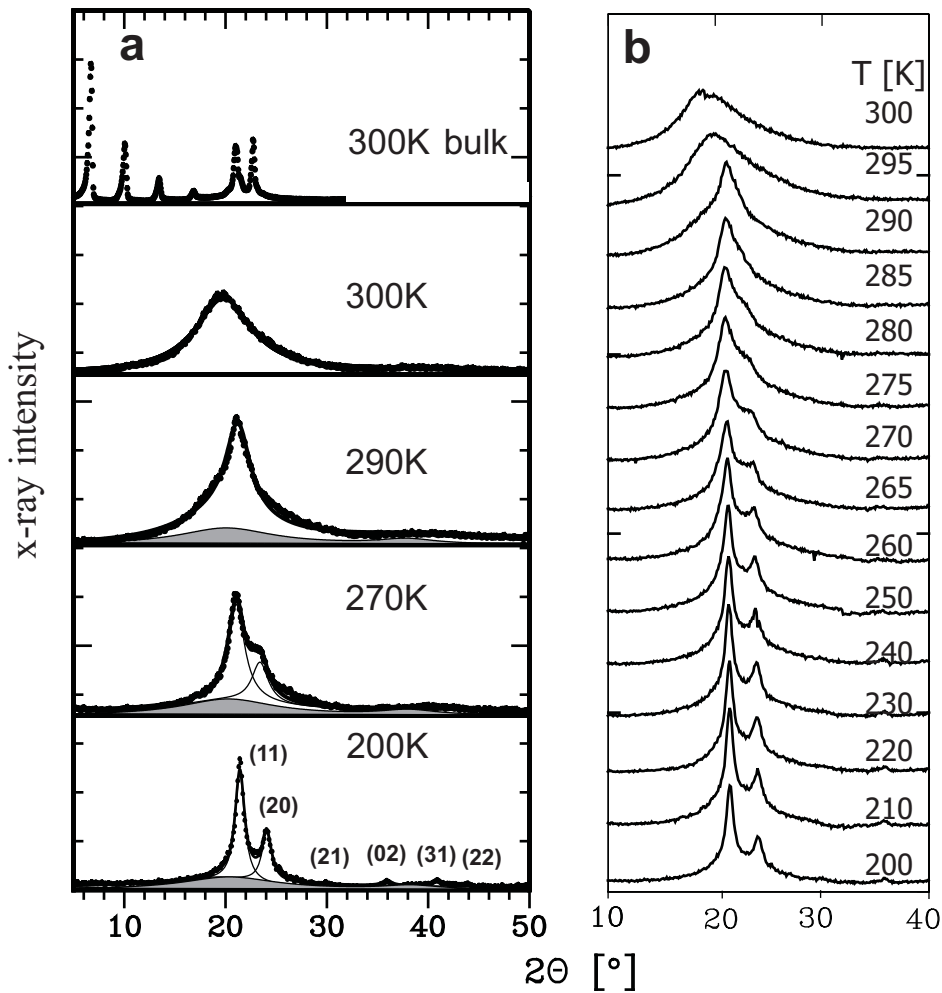} \caption{(a) Powder diffraction patterns
($\lambda $ = 1.542 {\AA}) of the R$_{I }$phase of bulk
C$_{19}$H$_{40}$ (top) and of the pore condensate. The four
examples shown represent the phases C and R$_{I}$, the coexistence
R$_{II}$-liquid and the liquid state. The reflections of phase C
are indexed on the basis of a two-dimensional rectangular mesh.
The solid line connecting the measured intensities (circles) is
the sum of the Lorentzian-type Bragg peaks plus an amorphous
background (gray shaded pattern) for the confined C and R$_{I}$
phase. In the case of the R$_{II}$-liquid state at T=290K an
additional scattering component with reduced intensity but with
the same overall shape as in the case of the pure confined liquid
at T=300K is necessary in order to arrive at a full fit of the
observed intensity. (b) Temperature dependent diffraction patterns
of the confined alkane taken while heating from 200K to 300K in 5K
steps.} \label{fig02}\end{figure}

For an inspection of the structure of the alkane in the pores,
diffraction patterns have been recorded on heating and cooling
between 200~K and 310~K using a Bragg-Brentano para-focussing
geometry\cite{Cullity1978}. The radiation comes from a rotating Cu
anode operating at 10kW and is monochromatized by reflection from
a graphite $(002)$ crystal ($\lambda=1.542\AA$). The diffraction
patterns cover scattering angles $2\Theta$ from $4^{\circ}$ to
$50^{\circ}$. The scattering background shows a broad maximum due
to the first peak in the structure factor of the amorphous
$SiO_{2}$ matrix. This background, which amounts to about 10$\%$
at its maximum value at 22 deg, has been subtracted, properly
accounting for the different x-ray absorption lengths. The
resulting patterns are shown in Fig. \ref{fig02}. The overall
appearance is totally different from the bulk reference pattern:
Stronger diffracted intensity is confined to scattering angles of
about 20deg. In particular the (00l) series of the bulk state has
disappeared. The pattern of the confined alkane at T=200K can be
derived from the bulk pattern by projecting the (hkl) Bragg peaks
onto the basal plane, allowing for a slight adjustment of the
in-plane lattice parameters $a$ and $b$. To arrive at a full fit
of the diffraction pattern, however, additional to the Bragg
peaks, a broad background indicating highly disordered structures
is necessary. From filling fraction dependent diffraction
experiments on e.g. Ar \cite{Huber1999} it is known that these
disordered structures stem from the first few monolayers of
molecules that are strongly adsorbed to the pore walls. Thus we
find a partitioning of the adsorbed material into a more or less
ordered and a highly disordered component. In the following we
shall focus on the first, more ordered component and denote it
"pore solid". The absence of any (00l) reflections in the
diffraction patterns means that the pore solid is an effectively
2D arrangement of molecules with random $z$-coordinates, while its
in-plane arrangement has changed little. Hence one basic geometric
ordering principle of the bulk crystalline alkane, i.e. the
layering, is suppressed.

\begin{figure}
\onefigure{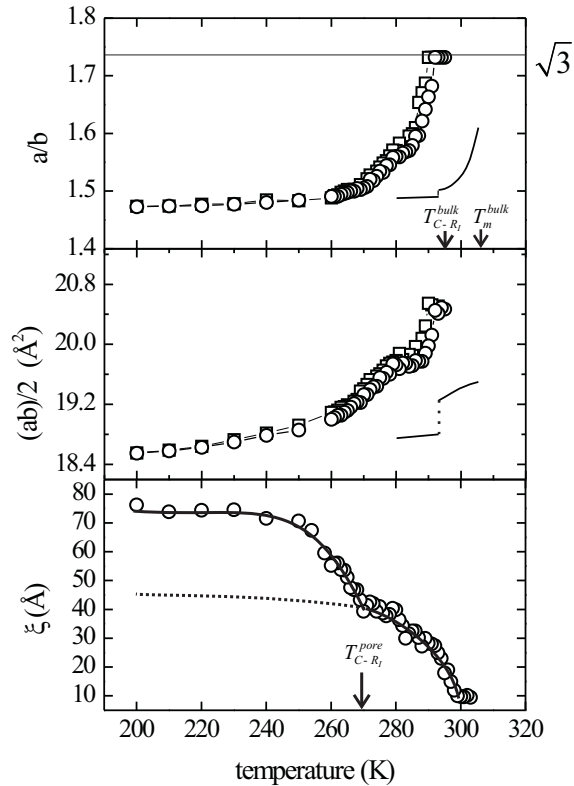} \caption{The temperature dependence of the
area per molecule in the lateral plane ( = pore cross section), of
the ratio of the in-plane lattice parameters a and b, and of the
coherence length $\xi$ as determined from Lorentzian fits while
cooling ($square$) and heating ($circ$). Results on the bulk
alkane in-plane parameters (\cite{Doucet1981}) are included for
comparison. The solid line in the ($\xi$,$T$)-plot is a guide for
the eye. The value of $\xi$ of the R$_{I}$ phase extrapolated
below T$_{C-R_{I}}$ is indicated by the dashed line.}
\label{fig03}
\end{figure}

An analysis of the coherence length $\xi$ based on the widths of
the diffraction peaks fitted with Lorentzians yields a continuous
increase of $\xi$ from about 10$\AA$ in the liquid phase to about
$\xi_{max}=75\AA$ at $T=200K$ - compare Fig.~\ref{fig03}. The
saturated value $\xi_{max}$ is close to the mean pore diameter of
$70\AA$ in Vycor; a result which can be explained only by an
arrangement of the molecules' long axes parallel to the pore axis.
A schematic view of the resulting structure of the pore solid in
respect to the pore geometry is depicted in Fig.~\ref{fig01}.
Despite the quenched lamellar ordering, the confined alkane
manages to order in a 2D close-packed arrangement. An observation
which is maybe not too surprising, if one considers the analogous
results on pore condensed Kr, Ar, N$_{2}$, and CO
\cite{Morishige2000Krypton,Huber1999}. These systems establish
spherical close packed structures, not only in the bulk state but
also in the confined state.

Does the $T$-evolution of the lateral arrangement in the pores
have any relation to the phase sequence of the bulk system?  In
order to explore this question, we consult the diffraction
patterns of fig. 2 and two quantities derived from these patterns
after fitting the diffraction peaks with Lorentzian profiles: the
area per molecule $ab$/2 in the lateral plane and the $a/b$ ratio.
Fig. 3 shows the $T$-dependence of these parameters, both in the
pores and in the bulk state. It is understood that one cannot
expect to observe sharp phase transitions in the pores of Vycor
glass. As in the case of the melting temperature mentioned above,
phase transition temperatures in the pores are usually lower than
those of the bulk \cite{Wallacher2001Rev}. The shift $\Delta T =
T_{\rm s}^{bulk}$-T$_{\rm s}^{\rm pore}$ is roughly proportional
to the inverse pore diameter. Therefore, the distribution of the
pore diameters in Vycor glass inevitably leads to a smearing of
the transition. Still, the behavior of the pore system is very
similar to that of the bulk system, provided that we allow for a
shift of the $T$-scale. The transition from phase C to the rotator
phase R$_{I}$ of the bulk solid is a first order transition, as
the discontinuities of $ab/2$ and $a/b$ indicate. In the pores one
observes at least clear changes of slope of the T-dependence of
these parameters. Another indication that this transition actually
occurs in the pores is supplied by the appearance of Bragg
reflections at higher scattering angles. Due to the partial
orientational disorder of the rotator phase,  which implies large
Debye-Waller factors, these Bragg reflections are absent in the
$T$-range of the R$_{I}$ phase. By contrast, they are clearly
exhibited in the low-$T$ C phase, comp. Fig.~\ref{fig02}, T=200K.

As can be seen in Fig.~\ref{fig03}, the a/b ratio of the confined
alkane's 2D lattice approaches and actually reaches the value of
$\surd $3, signaling the hexagonal metric of the rotator phase
R$_{II}$. The bulk solid, however, melts before entering this
phase. Thus, the R$_{II}$ phase is an extra feature of the pore
solid. The corresponding diffraction result is the one-peak
pattern at 290~K (cf. Fig.~\ref{fig02}). This pattern can be
decomposed into a sharper peak representing the R$_{II}$ pore
solid and a broader component representing the pore liquid. ( A
single-phase pattern of the pore liquid (at 300~K) is also shown
in Fig. 2.) Obviously the pore solid melts via an intermediate
state in which the liquid coexists with the R$_{II }$ phase,
although the R$_{II}$ phase is never observed in pure form. The
characteristic temperatures of the phase sequence C - R$_{I }$-
mixed state R$_{II}$/liquid~-~liquid are 269~K - 289~K - 294~K on
warming and 266 K -- 288 K -- 292 K on cooling, respectively. The
uncertainty is about 1~K.

Qualitatively, the quenching of lamellar ordering can be explained
in simple terms. As adsorption studies on planar and porous
substrates as well as studies on liquid crystals in porous glasses
have shown\cite{Rabe1991}, the interaction of alkanes and other
van-der-Waals systems with the host materials is stronger than the
intermolecular interaction. As monolayers, the n-alkanes lie flat
on these substrates, side by side, their long axis and the plane
of the zig-zag backbone parallel to the substrate. On a
cylindrical wall, rod-like molecules will have their long axis
parallel to the pore axis in order to profit most effectively from
their interaction with the substrate. The pore walls of porous
glasses are rough, since the matrix is an amorphous solid. On a
rough wall there is a hierarchy of adsorption sites. The molecules
select the deepest adsorption sites first. The positions of these
sites are uncorrelated. This leads to a randomization of the
$z$-coordinates of the molecules next to the walls. Obviously this
positional disorder is transferred to all other molecules in the
pore.

In respect to the larger $T$-width of the $R_{I}$ phase as well as
the appearance of the $R_{II}$ phase, one can argue in a similar
way. The orientation of the molecules next to the wall is dictated
by the wall; the C-C-plane is parallel to the wall and can no
longer adjust to the orientational pattern of the C or the R$_{I}$
phase. The interaction with the substrates breaks not only the
long-range translational order along $z$, but also the
orientational order. A rough wall may furthermore induce kink
defects which also destroy orientational order and are
characteristic of the orientational disordered rotator
phases\cite{Craievich1984}. One may speculate that the appearance
of the R$_{II}$ component owes to some part to such defects.

It is interesting to note, that complete quenching of lamellar
ordering has also been reported for natural waxes, e.g. bee waxes,
consisting of complicated mixtures of chain-like molecules, among
them alkanes\cite{Dorset1999}. There, one attributes this
phenomenon to the wide chain-length distribution, chain-branching
and partially to the chemical diversity of the system.

Our findings are also reminiscent of observations in systematic
studies of weakened interlayer interaction in alkanes, e.g. on
chain length mixing in alkanes\cite{Sirota1995} and the induced
weakening of this interaction by high pressure intercalation of
inert gases between the layers\cite{Sirota1994}. Always, one
observes an increased range of stability of the disordered phases,
particularly of the $R_{II}$ phase. Furthermore, in the complete
absence of interlayer interaction, as in surface crystallized
monolayers of alkanes floating on their own bulk liquid, only the
orientational disordered hexagonal phase is observed, even in
systems where this phase is not found in the bulk state, like
$C_{15}H_{32}$\cite{Wu1993}.

A closer inspection of the $T$-dependence reveals, that $\xi$
reaches the geometric limit dictated by the pore diameter d only
at about $T=240K$, well within the C phase. By contrast, at the
$R_I$ to C transition $\xi$ is only $\sim40\AA$ and even its
extrapolated value, below $T_{C-R_I}$, is only $\sim$ $45\AA$,
indicating that the R$_I$ as well as the R$_{II}$ pore condensed
phase is of short-range, rather than of long-range positional
order. Short-range order as well as a gradual evolution of
positional order are characteristic of 2D systems, 2D phase
transitions resp.\cite{Strandburg1988}. Along with the observed
complete loss of any positional order in the z-direction, this
encourages us to believe that the confinement leads here to a
reduced dimensionality of the system, rendering it 2D-like. In
fact, the $\xi$ behavior for the confined alkane agrees with the
one found for phase transitions of rod-like molecules in weakly
coupled 2D layers of smectic liquid crystals, which could be
described by 2D isotropic-hexatic-crystalline phase transitions
\cite{Pindak1981}. Thus, we propose a mapping of the
disorder-order transitions of the confined alkane on 2D models,
e.g. models based on the anisotropic planar rotator as suggested
for order-disorder transitions in alkane
monolayers\cite{Wurger1999} or models with an XY-like primary
order parameter, where this order parameter should be coupled to
the zig-zag backbone orientation, and the interaction should be of
a quadrupolar type. These models, enriched by external fields
accounting for the random disorder introduced by the matrix should
give deeper insights into the observed behavior.

In summary, the confined alkane manages to form 2D close-packed
structures in the pores of Vycor despite the drastic geometric
restrictions and while compromising one basic ordering principle
of the bulk alkane, i.e. the lamellar ordering.  Obviously the
randomization of the z-positions of the molecules does not
suppress the orientational order-disorder transitions known from
the bulk alkanes. On the other hand the confinement in the pores
favors the disordered phases, as documented by the downward shift
of the melting temperature, the appearance of the true rotator
phase R$_{II}$ and the increase in the temperature range of the
partially disordered phase R$_{I }$ at the expense of the
crystalline phase~C. As witnessed by the behavior of the coherence
length of the evolving positional ordering and the observed
analogies with alkanes with weakened interlayer interaction, the
complete loss of any ordering along the z-direction renders the
system 2D-like. Thus, the alkane confined within Vycor constitutes
a fine example of how nanoconfinement leads to new structures and
different phase behavior, here, in fact, to a reduction of the
effective dimensionality of the system, not towards 1D, as one
might have thought, but towards 2D.

\acknowledgements We would like to thank Prof. M. Deutsch
(Bar-Ilan University) for helpful suggestions. This work has been
supported by the Deutsche Forschungsgemeinschaft (SFB277).


\end{document}